\documentclass[aps,preprint,showpacs]{revtex4-1}

\usepackage{amsmath,amssymb}
\usepackage{bm}
\usepackage[dvips]{graphicx}

\def \A{\mathcal{A}}
\def \e{\varepsilon}
\def \w{\omega}

\begin{document}
\title{High momentum components in the nuclear symmetry energy}

\author{Arianna Carbone and Artur Polls}
\affiliation{Departament d'Estructura i Constituents de la Mat\`eria and Institut de
Ci\`{e}ncies del Cosmos, Universitat de Barcelona, Avda. Diagonal 647, E-08028 Barcelona, Spain,EU}

\author{Arnau Rios}
\affiliation{Department of Physics, Faculty of Engineering and Physical Sciences, University of Surrey,
Guildford, Surrey GU2 7XH, United Kingdom,EU}

\begin{abstract}
The short-range and tensor correlations associated to realistic nucleon-nucleon interactions
induce a population of high-momentum components in the many-body nuclear
wave function. We study the impact of such
high-momentum components on bulk observables associated to isospin asymmetric matter.
The kinetic part of
the symmetry energy is strongly reduced by correlations when compared to the non-interacting case.
The origin of this behavior is elucidated
using realistic interactions with different short-range and tensor structures.
\end{abstract}


\maketitle

The existence of high-momentum components in the nuclear many-body wave function is a well-established property 
both from an experimental \cite{rohe2004,subedi2008} and a theoretical points of view 
\cite{benhar1994a,dickhoff2004a}. Short-range correlated pairs have been studied in detail 
at electron scattering facilities \cite{benhar2008a}. Two-nucleon knock-out reactions
have identified the predominance of isospin $I=0$ correlated pairs \cite{subedi2008}, 
an effect that has been 
related to the tensor component of the nucleon-nucleon (NN) interaction \cite{schiavilla2007,alvioli2008}.\\
\indent The enhanced effect of correlations on neutron-proton (np) pairs with respect to neutron-neutron (nn) 
pairs suggests that correlations can be increased (or even tuned) in isospin asymmetric systems, 
where a different number of np and nn  
pairs exist. Our microscopic many-body calculations indicate that the dependence of short-range and tensor correlations on the 
isospin asymmetry of the system, $\alpha=(N-Z)/(N+Z)$, is 
well understood from general theoretical principles \cite{frick2005}. 
One-body occupations, for instance, follow a systematic trend: neutrons become less depleted as 
the system becomes more neutron-rich, while the proton depletion increases \cite{frick2005,rios2009a}. \\
\indent Here, rather than looking at microscopic properties, we want to quantify the effect that NN correlations 
have on the bulk properties of isospin asymmetric systems. 
By analysing the energy of symmetric, asymmetric and neutron matter obtained within a realistic 
many-body approach, we will draw conclusions on the impact of correlations on asymmetric systems. We 
will focus our attention on the symmetry energy, which characterises 
the properties of isospin-rich nuclei as well as neutron stars \cite{Steiner2005}.\\
\indent To quantify the effect of correlations in a meaningful way, we use a many-body approximation 
that includes consistently short-range and tensor correlations. 
The ladder approximation, implemented within the
self-consistent Green's functions (SCGF) approach, provides a microscopic description of 
these effects via a fully dressed propagation of nucleons in nuclear matter 
\cite{dickhoff2004a}. This is achieved
by (a) computing the scattering of particles via a $T$-matrix (or effective interaction) in the medium, 
(b) extracting a self-energy out of the effective interaction and (c) using Dyson's equation to build
two-body propagators which are subsequently inserted in the scattering equation \cite{frick2005,Dieperink2003}. 
To solve the close set of equations, an iterative numerical procedure is a must. 
Recent advances have allowed implementations
both at zero \cite{soma2008} and at finite temperature \cite{rios2008c} using fully realistic NN 
interactions. 
We will focus here in calculations based on 
two-body forces only, including partial waves up to $J=4$ ($J=8$) for the dispersive (Hartree-Fock) contributions. 
\\
\indent 
Fully self-consistent ladder calculations in isospin asymmetric nuclear matter are scarce \cite{frick2005,rios2009a}. 
Previous results have mostly highlighted the different role of correlation in either symmetric or pure neutron matter 
\cite{Dieperink2003}. We will analyse here the asymmetry dependence of the different components of the energy. 
More specifically, we want to look into the kinetic component of the energy. The latter has been identified recently 
as a particularly good indicator of correlations in the nuclear ground state \cite{xu2011,vidana2011}. 
We will confirm these indications using SCGF techniques.\\
\indent The bulk properties of nuclear and neutron matter are obtained within the SCGF approach via 
the Galitskii-Migdal-Koltun sum-rule:
\begin{align}
\frac{E}{A} = \frac{\nu}{\rho} \int \frac{\textrm{d}^3 k}{(2 \pi)^3} \int \frac{\textrm{d} \w}{2 \pi} 
\frac{1}{2} \left\{ \frac{k^2}{2m} + \omega \right\} \A(k,\w) f(\w) \, ,
\label{eq:GMK}
\end{align}
where $\nu=4 \, (2)$ is the degeneracy of nuclear (neutron) matter, $\rho$ is the total density and $f(\omega) = \left
[ 1+\exp{(\omega - \mu)/T} \right]^{-1}$ is a Fermi-Dirac distribution. 
Roughly speaking, the one-body spectral function, $\A(k,\w)$, 
represents the probability of knocking out or 
adding a particle with a given single-particle momentum, $k$, and energy, $\omega$. 
The spectral function also gives access to all the one-body operators of the system \cite{dickhoff2005}.
For instance, the momentum distribution, $n(k)$, is obtained by convoluting the spectral function with a Fermi-Dirac factor:
\begin{align}
n(k) = \int \frac{\textrm{d} \w}{2 \pi} \A(k,\w) f(\w) \, .
\label{eq:momdis}
\end{align}
Correlations beyond the mean-field approximation 
have a particularly clear manifestation in the momentum distribution \cite{rios2009a}. 
A sizeable depletion appears below the Fermi sea, while high-momentum components are populated \cite{Muther2000}.
\begin{figure}
\begin{center}
       \includegraphics[width=0.95\linewidth]{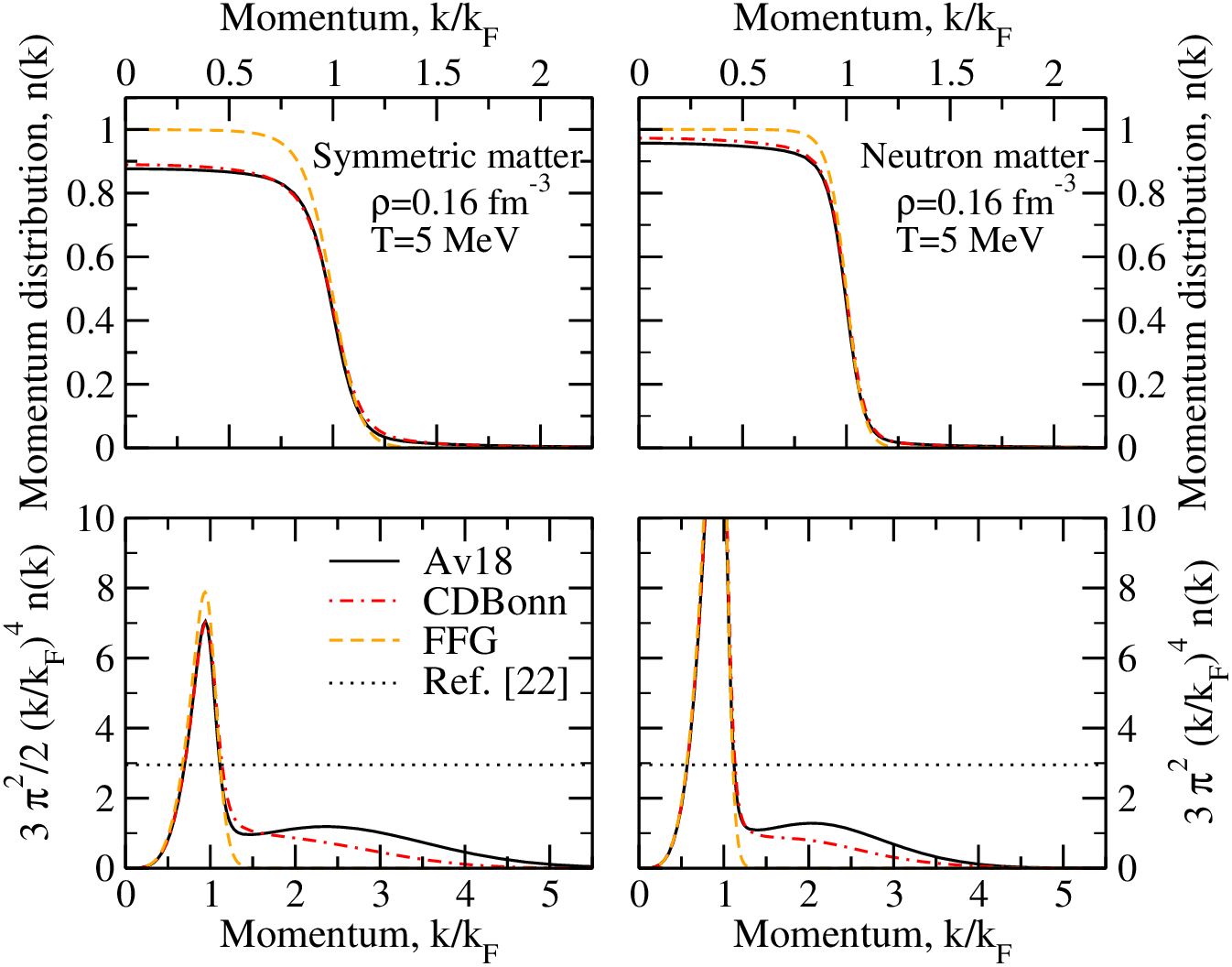}
       \caption{(Color online) Upper left (right) panels: momentum distribution of symmetric nuclear
        (neutron) matter obtained with the SCGF approximation for Av18 (full lines) and CDBonn (dash-dotted lined) 	
        and for the FFG (dashed lines). 
        Lower panels: the momentum distribution times $k^4/k_F^4$, a quantity directly connected to the kinetic energy. 
        All the results are computed at $\rho=0.16$ fm $^{-3}$ and $T=5$ MeV. The dotted line shows the high 
        momentum constant of Ref.~\cite{drut2011}}
       \label{fig:momdis}
\end{center}
\end{figure}
\indent To illustrate this point, we show in the upper left (right) panel of Fig.~\ref{fig:momdis} the momentum 
distribution for symmetric nuclear (pure neutron) matter. 
The results obtained within the SCGF method for the Argonne v18 (Av18) \cite{wiringa1995} and the CDBonn 
 \cite{machleidt1995} interactions (solid and dash-dotted lines, respectively) are compared to the momentum 
 distribution of the Free Fermi Gas (FFG) in the
same conditions (dashed lines). The FFG is used here as a benchmark for thermal effects.
A common feature of the SCGF and the FFG results is the 
softening of the distribution around the Fermi surface, $k=k_F$, associated to the finite temperature of 
$T=5$ MeV. Calculations are performed at this temperature to avoid 
the pairing instability \cite{alm1996}. 
The density, $\rho=0.16$ fm$^{-3}$ does not correspond
to the saturation point of any of the two potentials. For Av18 (CDBonn) within the SCGF approach at $T=5$ MeV, the saturation density is at $\rho=0.19$ ($0.32$) fm$^{-3}$. Three-body forces (3BFs) should improve the saturation properties without strongly affecting isovector properties \cite{soma2008}.
\\
\indent Correlation effects in the momentum distribution
are substantially different in symmetric nuclear matter (SNM) and in pure neutron matter (PNM) 
\cite{rios2009a}. The presence of the $S-D$ tensor interaction in the isospin-saturated system induces 
a larger amount of correlations. 
As a consequence, the Fermi surface is quite more depleted for SNM than for PNM (compare
the upper left and right panels in Fig.~\ref{fig:momdis}).
Typical values for these depletions are obtained from the occupation at zero momentum:
$n(0) \sim 0.87 $ for SNM and $n(0) \sim 0.96 $ for PNM. 
Because the momentum distribution is normalised to the total density,
the high momentum components are also rather different for 
both systems at the same density. 
One can summarise these differences by looking at the integrated strength over different regions
of momenta:
\begin{align}
\phi_m(k_i,k_f) =  \frac{\nu}{2 \pi^2 \rho} \int_{k_i}^{k_f} \textrm{d} k \, k^{m} n(k) \, .
\label{eq:strength}
\end{align}
The integral with $m=2$ represents the 
fractional contribution of a given momentum region to the total density. Similarly, the integral with $m=4$ is  
related to the total kinetic energy of the system.
\begin{table}[t!]
\begin{center}
\scalebox{0.8}{
	\begin{tabular}{l|c c|c c c|c c c}
	    $\phantom{a}$ & $\phantom{a}$ & $\phantom{a}$ & 
	    $\phantom{a}$ & SNM & $\phantom{a}$ &  $\phantom{a}$ & PNM & $\phantom{a}$  \\ \hline
		$\phantom{a}$ & $k_i$ & $k_f$ & $\phi_2$ & $K/A$& $E/A$ & 
		$\phi_2$ & $K/A$ & $E/A$ \\ \hline
	    CDB         & $0$    & $k_F$    & 0.762 & 15.8 & -12.6 & 0.869 & 28.9 & 10.1 \\
		$\phantom{a}$ & $k_F$  & $2k_F$   & 0.211 & 12.0 & -1.54 & 0.121 & 9.87 & 3.78  \\
		$\phantom{a}$ & $2k_F$ & $\infty$ & 0.027 & 6.95 & -0.59 & 0.010 & 3.78 & 0.28 \\
		$\phantom{a}$ & $0$    & $\infty$ & 1.00  & 34.7 & -14.7 & 1.00  & 42.5 & 14.1 \\
		\hline
		 Av18        & $0$    & $k_F$    & 0.755 & 15.6 & -7.65 & 0.863 & 28.7 & 11.6 \\
		$\phantom{a}$ & $k_F$  & $2k_F$   & 0.194 & 11.4 & -0.997& 0.119 & 10.3 & 3.24  \\
		$\phantom{a}$ & $2k_F$ & $\infty$ & 0.051 & 14.5 & -1.29 & 0.018 & 7.16 & 0.32\\
		$\phantom{a}$ & $0$    & $\infty$ & 1.00 &  41.5 & -9.94 & 1.00 & 46.2 & 15.2\\
		\hline
		FFG         & $0$    & $k_F$    & 0.861 & 17.7 & 17.7 & 0.912 & 30.4 & 30.4 \\
		$\phantom{a}$ & $k_F$  & $2k_F$   & 0.139 & 6.00 & 6.00 & 0.089 & 5.75 & 5.75 \\
		$\phantom{a}$ & $2k_F$ & $\infty$ & 0.00  & 0.00 &  0.00& 0.00  & 0.00 &  0.00\\
		$\phantom{a}$ & $0$    & $\infty$ & 1.00  & 23.7 & 23.7 & 1.00  & 36.2 & 36.2 \\
		\hline
	\end{tabular}
}
\end{center}
	\caption{Contributions of different momentum regions to the total density (columns 3 and 6),
 kinetic (4 and 7, in MeV) and total energies (5 and 8, in MeV) for SNM (columns 3, 4 and 5) and PNM (columns
6,7 and 8) with different NN interactions. The FFG case is also included. All the results are computed at
 $\rho=0.16$ fm$^{-3}$ and $T=5$ MeV.}
	\label{table:snm}
\end{table}

In Table~\ref{table:snm}, we give the integrated strengths of SNM and PNM 
at $\rho=0.16$ fm$^{\textrm{-3}}$ and $T=5$ MeV for
the Av18 and the CDBonn  interactions. 
As expected, in SNM there is a substantial depletion of states below the Fermi surface, 
\emph{i.e.} only $\sim 75 \, \%$ of the strength is
in the region between $0$ and $k_F$. Part of the depletion has a thermal origin, and the 
comparison with the FFG in the same momentum region
 suggests that between $1/2$ and  $2/3$ of the integrated depletion  
comes from the softening of the Fermi surface due to
the finite temperature. The effect of correlations is also important beyond the Fermi surface: for SNM 
(PNM) there is still a $3-5 \, \%$ ($1-2 \, \%$) of the strength in the region  $k > 2k_F$. 
The energy per particle is also affected by short-range correlations \cite{Dewulf2003}. In particular,
the kinetic energy, 
\begin{align}
\frac{K}{A} = \frac{\nu}{\rho} \int \frac{\textrm{d}^3 k}{(2 \pi)^3}  \frac{k^2}{2m} n(k) \, ,
\label{eq:kinetic}
\end{align}
increases with respect to the FFG due to the population of high-momentum components \cite{Muther2000}. 
We present the  contributions of the different momentum regions  to the kinetic energy of nuclear and 
neutron matter in columns 4 and 7 of Table~\ref{table:snm}. 
First of all, let us note that the total integrated values of the correlated kinetic energies 
 are larger than those of the FFG. 
For SNM, these are $\sim 11$ and $\sim 17$ MeV larger for CDBonn and Av18, 
respectively. The difference between the two potentials is expected, since Av18 is a hard
interaction and has a stronger tensor component than CDBonn. In PNM the total kinetic 
energy is $6$ and $10$ MeV higher, respectively, for both interactions. 
This emphasises again the idea that correlation effects play a smaller role in PNM than in SNM.\\
\indent Let us stress once more the importance of components beyond $k_F$ in the kinetic energy. 
For SNM, these amount
to more than $50 \, \%$ of the total, while in PNM they account for more than $25 \, \%$. 
In contrast, for the FFG at this temperature, the contribution of states above
$k_F$ is less than $25 \, \%$ ($15 \, \%$) for SNM (PNM). 
The FFG strength above the Fermi surface is due to thermal effects, which tend to be
localised within a small region around $k_F$ for low temperatures. As a consequence, there 
are no contributions to the FFG energy in the region of $k>2k_F$. The contributions in this 
region in the interacting case can be entirely attributed to NN
correlation effects. For a hard interaction like Av18, the contribution beyond $2k_F$ in SNM 
is even larger than that between the Fermi surface and $2k_F$. \\
\indent A visual representation of the contributions of different momentum regions to the kinetic
energy is obtained by looking directly at the integrand in 
the formula for the kinetic energy, $k^4 n(k)$. The lower panels of Fig.~\ref{fig:momdis} show this 
quantity for SNM (left) and PNM (right). 
Compared to the FFG, one finds that the integrands for SNM and PNM have substantial 
contributions for $k > k_F$. In both cases, the integrand extends to very high 
momenta, up to $4-5 k_F$. 
With the present normalization, the high momentum components of both systems
are similar in size, but they extend to higher momenta for SNM than for PNM. 
As already discussed, the kinetic energy of SNM is higher than that of PNM 
with respect to the FFG. This will have a strong impact on the kinetic component of the symmetry energy. \\
\indent If SNM or PNM were in the unitary regime, the integrands shown in the lower panels of Fig.~\ref{fig:momdis}
would tend to the so-called contact constant, $C/(Nk_F)$, at very high momenta \cite{drut2011}. 
For a unitary gas, the momentum distribution in the $k \to \infty$ region would therefore decay as $k^{-4}$.
SNM and PNM are not unitary gases, though, and
their kinetic energy integrands are not constant at high momentum. Instead, these functions
have a peak at $k_F$ and then decreases sharply (at $T=0$ there would be
a discontinuity at $k_F$). In the region $1.5-3 k_F$, $k^4 n(k)$ levels off (signalling a plausible
$n(k) \sim k^{-4}$ scaling in this region) and then decays softly as $k$ increases. 
The presence of a secondary maximum in the Av18 results explains why the kinetic energy has such a 
large contribution in the $k>2k_F$ region. To ensure the correct high momentum limit, 
the integrands have been normalised by the constant $(3 \pi^2)/(\nu)$, 
\emph{i.e.} the normalizations in SNM and PNM differ by a factor of $2$.
With this, the magnitude and shape of the high momentum components look quite similar 
for the two systems and interactions considered in Fig.~\ref{fig:momdis}. 
For SNM (PNM) under these
conditions, we have $T/\e_F \sim 0.14 \, (0.08)$ and we find that, in both cases, the maximum
value of the integrand is below $2$. In the same conditions, quantum Monte-Carlo calculations of
a unitary gas suggest a value of $C/(Nk_F) \sim 3$ \cite{drut2011}. 
\begin{figure}
\begin{center}
       \includegraphics[width=0.42\linewidth]{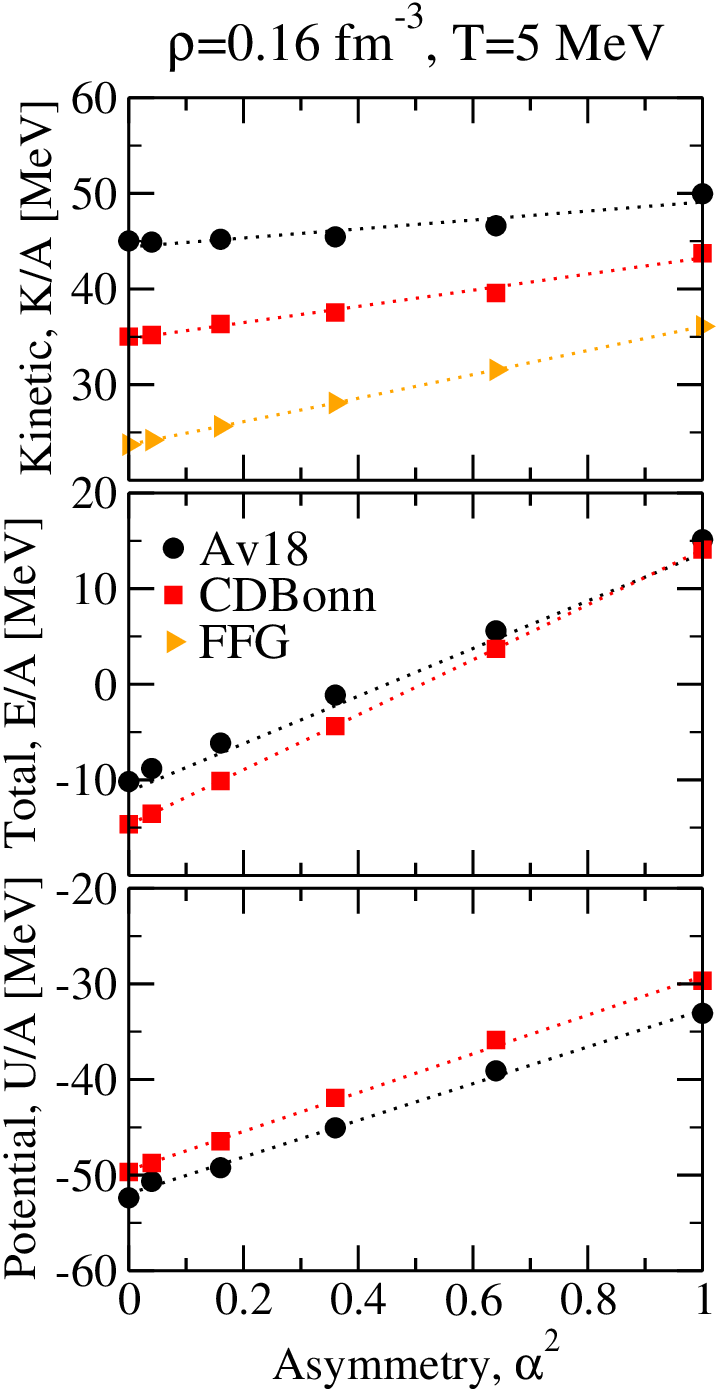}
       \caption{(Color online) Isospin asymmetry dependence of different components of the energy for 
       the CDBonn (circles) and Av18 (squares) potentials. The upper, central and lower panels correspond 
       to the kinetic, total and potential energies, respectively. The triangles of the upper panel give 
       the energy of the Free Fermi Gas in the same conditions, $\rho=0.16$ fm$^{-3}$ and $T=5$ MeV. 
       Dotted lines are linear regressions to guide the eye.}
       \label{fig:easym}
\end{center}
\end{figure}

\begin{figure}
\begin{center}
       \includegraphics[width=0.9\linewidth]{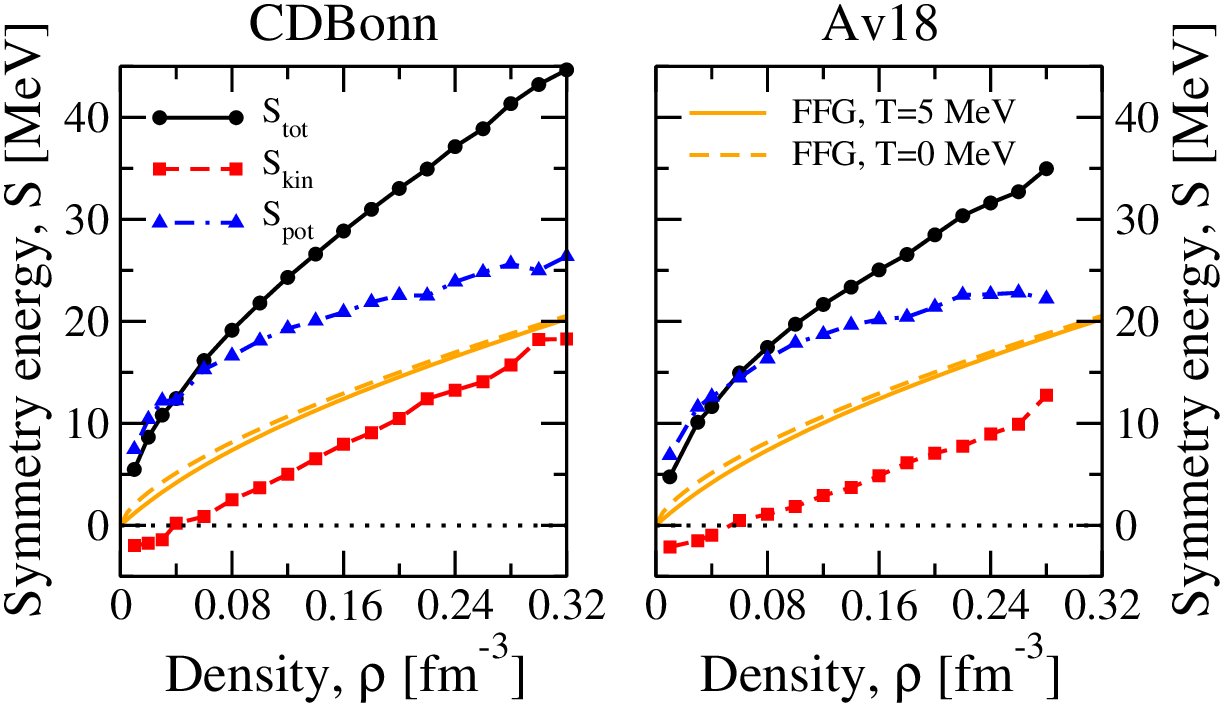}
       \caption{(Color online) Components of the symmetry energy for the CDBonn (left panel) 
       and Av18 (right panel) potentials at T=5 MeV. Circles, squares and triangles represent the total, 
       kinetic and potential contributions, respectively. The continuous (dashed) lines correspond to 
       the FFG symmetry energy at T=5 (T=0) MeV.}
       \label{fig:esym_rho}
\end{center}
\end{figure}

\indent To compute the symmetry energy (and its kinetic and potential components), one generally 
resorts to the parabolic formula, \emph{i.e.} one assumes that the energy per particle 
(or any of its  components) has a quadratic dependence on asymmetry,
\begin{align}
	\frac{E}{A}(\rho,\alpha)= 	\frac{E}{A}(\rho,0) + S(\rho) \alpha^2 \, .
\end{align}
This immediately yields that the symmetry energy, $S(\rho)$, is given by the difference of PNM and SNM energies:
\begin{align}
	S(\rho) = \frac{E}{A}(\rho,1) - \frac{E}{A}(\rho,0) \, .
	\label{eq:diff}
\end{align}
\indent The SCGF approach can be generalised to isospin asymmetric 
systems \cite{frick2005}. We can thus validate the parabolic assumption by performing 
calculations at different $\alpha$'s and looking at the explicit dependence of the energy on
isospin asymmetry. We have performed this check
for the three components
of the energy (kinetic, potential and total) and two representative potentials, CDBonn (circles) and Av18 
(squares),
at $\rho=0.16$ fm$^{\textrm{-3}}$ and $T=5$ MeV. The results are displayed in the three panels of 
Fig.~\ref{fig:easym}, as a function of $\alpha^2$. 
Note that the vertical extent of all three panels is the same. 
In general, the three components seem to have a well-defined parabolic
dependence on $\alpha$. To stress this point, we show the results of a linear regression fit (dotted lines)
on top of the calculated data.
The slope of the linear regression reduces to the different components (kinetic, potential and total) 
of the symmetry energy if the parabolic approximation holds exactly. We have found a very good 
agreement (generally within less than $0.5$ MeV) between
the slopes and the values obtained using the differences of Eq.~(\ref{eq:diff}).\\
\indent In the upper panel of Fig.~\ref{fig:easym}, we compare the kinetic energy of the two potentials 
to that of the corresponding FFG (triangles).
As expected, the correlated kinetic energy is larger than the FFG at all asymmetries. A hard 
interaction (Av18) leads to a substantially larger kinetic energy than a soft one (CDBonn). 
Moreover, the isospin dependence of both forces is different (and different than the FFG). 
While the kinetic energy of Av18 in SNM ($\alpha=0$) is 
$K/A \sim 42$ MeV and that of PNM is $\sim 46$ MeV, for CDBonn
these two quantities are $35$ and $43$ MeV, respectively. 
In other words, the difference between the kinetic energies of PNM and SNM is smaller for Av18 than for 
CDBonn. Correspondingly, both of these differences are smaller than that associated to the FFG.\\
\indent The small value of the kinetic symmetry energy in a correlated approach is one of the major conclusions
of this paper. One can understand the origin of this behaviour from the following reasoning.
The tensor component of the NN force, acting on SNM, induces large correlations and produces an 
important renormalization of the kinetic energy with respect to the FFG.
The absence  of this component in PNM reduces the relative enhancement of the kinetic energy. 
Consequently, the difference in total kinetic energies is smaller for the correlated case than for the 
FFG value. In turn, this implies that the kinetic symmetry energy is reduced. 
Within this picture, tensor correlations, and the renormalization they induce in the kinetic energy,
seem to be the main  responsible for the small values of the kinetic symmetry energy.\\
\indent The asymmetry dependence of the total energy per particle is driven by a competition between the kinetic and the 
potential terms. 
The size of both contributions is density dependent but, at $\rho=0.16$ fm$^{-3}$, 
the potential term largely dominates the isospin dependence. This can be 
directly seen in Fig.~\ref{fig:easym}: the difference
between the PNM and SNM potential energies is of the order of $20$ MeV, while for the 
kinetic term the differences are below $10$ MeV.
\begin{table}[t!]
\begin{center}
\scalebox{0.8}{
	\begin{tabular}{c| c c c c}
		$\phantom{a}$ & $S_{\textrm{tot}}$ & $S_{\textrm{kin}}$ & $S_{\textrm{pot}}$ & $L$ \\
		\hline
		Av18 & 25.1 & 4.9 & 20.2 & 37.7 \\
		Nij1 & 27.4 & 4.6 & 22.8 & 48.5 \\
	  CDBonn & 28.8 & 7.9 & 20.9 & 52.6 \\
		N3LO & 29.7 & 7.2 & 22.4 & 55.2 \\
	\end{tabular}
}
\end{center}
	\caption{Total, kinetic and potential contributions in MeV to the symmetry 
energy at $\rho=0.16$ fm$^{\textrm{-3}}$ and $T=5$ MeV for different NN interactions.
 The last column gives the  $L$ coefficient in MeV, related to the density dependence of $S_{\textrm{tot}}$
 in the same conditions.}
	\label{table2}
\end{table}
\indent The values of the different components of the symmetry energy obtained in the parabolic approximation
at $T=5$ MeV are given in Table~\ref{table2}. In addition to Av18 and CDBonn, 
we consider two other  representative NN interactions: the somewhat hard Nijmegen 1 \cite{stoks1994} and 
the very soft N3LO with a $\Lambda=500$ MeV cut-off \cite{entem2003}. 
Note that the calculations have 
been performed at $\rho=0.16$ fm$^{-3}$, which is not the saturation density of any of these potentials.
Based on the comparison with the FFG (see Fig.~\ref{fig:esym_rho}), 
the effect of finite temperature in the symmetry energy should be rather small. 
At the density we are considering, the symmetry energy of the FFG increases from $12.4$ 
MeV at zero temperature to $13$ MeV at $T=5$ MeV (see below for further discussion). \\
\indent The total symmetry energies predicted by the SCGF approach range
between $25$ and $30$ MeV, just below the 
currently accepted $\sim 32$ MeV value \cite{Tsang2009a}. 
The inclusion of 3BFs would bring the SCGF results closer to experiment. 
Although we cannot perform calculations with 3BFs at the moment,
we can estimate their importance from existing Brueckner--Hartree--Fock (BHF) calculations \cite{vidana2009}.
Around $\rho=0.16$ fm$^{-3}$, 3BFs tend to increase the symmetry energy by $3-4$ MeV \cite{vidana2009}. 
\\
\indent The symmetry energies provided
by SCGF calculations tend to be smaller than the BHF ones with the same
two-body NN force \cite{vidana2009}. 
The origin of this difference can be understood as follows. In general, the propagation 
of holes and the dressing of the intermediate propagators in the ladder equation has an 
overall repulsive effect in the total energy of the system with respect to the BHF values \cite{Muther2000}.
This repulsive effect is larger for SNM, where correlations are more substantial, than for PNM. 
The difference between energies is therefore reduced and the SCGF symmetry energy becomes smaller than the BHF one. 
Since this repulsive effect increases with 
density, the slope of the symmetry energy as a function of the density is also expected to decrease. 
At this point, it is worth mentioning that the BHF symmetry energy obtained with the Av18
interaction, at the same density and temperature as in Table~\ref{table2}, is $S_{\textrm{tot}}=28.4$ MeV,
\emph{i.e.} about $\sim 3$ MeV higher than the SCGF result.
Similar conclusions have been discussed previously in the context of SCGF calculations \cite{Dieperink2003}
and of extensions of BHF theory inspired by Green's functions
 \cite{Hassaneen2008}.
\\ 
\indent As already explained, the kinetic symmetry energy (column 3 of Table~\ref{table2}) is small. 
In most cases it lies between $4-8$ MeV, \emph{i.e.} well below the corresponding FFG value of $S_{\textrm{kin}} = 12$ MeV. 
Such a small value is compatible with recent observations obtained either with a simplified phenomenological model, 
as in Ref.~\cite{xu2011}, or with a more sophisticated many-body method based on the 
Brueckner-Hartree-Fock (BHF) theory, as in Ref.~\cite{vidana2011}.
Similarly, Fermi-Hypernetted-Chain (FHNC) calculations with realistic two and three-body interactions
also yield relatively small kinetic symmetry energies \cite{lovato2011,lovato2011b}.
One can therefore conclude that this effect is caused by NN correlations in the many-body wave-function.\\
\indent For illustrative purposes, we also show, in the last column of Table~\ref{table2}, the value of the $L$ 
parameter, $L = 3 \rho \textrm{d} S/\textrm{d} \rho$, obtained in the same conditions. Most SCGF results fall 
below the currently preferred value of $L \sim 50-60$ MeV \cite{Tsang2009a}. 
The important corrections induced by 3BFs will improve this result \cite{soma2008,vidana2009}.
Compared to BHF results with the same interactions, we find a decrease of the slope of the symmetry energy around
$\rho=0.16$ fm$^{-3}$, in qualitative accordance to Ref.~\cite{Dieperink2003}.
As an example, for Av18 at the same temperature, the BHF approach provides $L= 47.5$ MeV.\\
\indent Our results show an almost linear correlation between the values of the symmetry energy, $S$,
and the slope parameter, $L$. Lower symmetry energies, generally associated to harder interactions, correspond
to lower slopes. Similarly, softer interactions seem to induce higher symmetry energies and slopes. 
This confirms the trends observed in
phenomenological mean-field calculations from a purely microscopic perspective \cite{vidana2009,centelles2009}. 
In other words, one can think of the correlation between $S$ and $L$ as a general property of isospin asymmetric
systems, rather than as an artifact due to the fitting procedures of mean-field parametrizations. 
We illustrate the density dependence of our results in Fig.~\ref{fig:esym_rho}, where we show
the total, kinetic and potential components of the symmetry energy obtained with CDBonn (left panel)
and Av18 (right panel) at $T=5$ MeV. The symmetry energy and its components grow steadily with density in the
density range considered here. The potential component is always larger in absolute
value than the kinetic one, thus dominating the contribution to $S$.
The difference in the density dependence of the kinetic and potential contributions is
subtle. While $S_{\textrm{kin}}$ grows almost linearly with $\rho$, 
$S_{\textrm{pot}}$ seems to have a milder density dependence. 
In absolute terms, however, both contributions have a similar importance for $L$ around saturation density.
Similar conclusions hold for other NN interactions, not shown here for simplicity.\\
\indent It is interesting to note that the kinetic symmetry energy becomes negative below a 
density of about $0.04-0.08$ fm$^{-3}$. At such low densities, the effect of thermal
correlations is expected to be important, so one might be tempted to attribute this 
anti-intuitive behaviour to finite temperature correlations. 
To pin down the importance of thermal effects, we also show in both panels of 
Fig.~\ref{fig:esym_rho}, the symmetry energy
of the FFG at $T=0$ (dashed line) and $T=5$ (solid) MeV. The differences are extremely small: 
the symmetry energy decreases by less than $1$ MeV when going
from zero to a temperature of $5$ MeV in the whole density range. 
The small contribution of temperature on the symmetry energy is caused by the relatively
similar thermal corrections of SNM and PNM \cite{rios2008c}. When taking the difference of both energies,
one eliminates practically the temperature dependence.
As a matter of fact, in both the classical (low density) and degenerate (high density) limits, 
the thermal effects on the symmetry energy disappear exactly. 
For the SCGF results, this suggests that 
negative kinetic symmetry energies at low densities are in fact not a thermal, but a correlation-dominated effect. \\
\indent A negative kinetic symmetry energy has already been found (although not explicitly
mentioned) in Ref.~\cite{xu2011}, where a simplified model for high-momentum components
was proposed by assuming a given shape of $n(k)$. 
The model depends on a single, density-independent parameter, $a$, which accounts for the
depletion of states below $k_F$. The original values of $a$ seem to be somewhat extreme when
compared with depletions obtained in realistic approaches. As a matter of fact, tuning  $a$ 
to reproduce our momentum distributions eliminates the negative values of $S_{\textrm{kin}}$. 
With zero-temperature FHNC calculations, one also obtains negative kinetic symmetry
energies at low densities \cite{lovato2011b}. \\
\indent We would like to stress that, in principle, negative values of
$S_{\textrm{kin}}$, unlike negative values of $S_{\textrm{tot}}$, are not associated to a thermodynamical 
instability \cite{margueron2003}. In general, our results suggest that one needs to take
into account high momentum components to get realistic values of the kinetic symmetry energy. 
Whether or not negative values of $S_{\textrm{kin}}$ (or their underlying cause, high momentum components)
have an impact on either neutron star physics or transport simulations remains to be seen. \\
\indent 
We have focused here on the kinetic component of the symmetry energy because of its close and transparent
relation to short-range and tensor correlations. The potential energy contribution is equally
important for isospin physics. However, due to the presence of the mean-field contribution, high momentum 
correlations will be harder to disentangle in the potential symmetry energy. We expect both its value 
at saturation as well as its density dependence
to be different than those predicted by mean-field theories. 
\\
\indent In summary, we have studied the isospin asymmetry dependence of the bulk properties of
nuclear matter within the SCGF approach. We have confirmed the quadratic dependences 
on isospin asymmetry of the total, potential and kinetic energy. 
We have highlighted the effect of NN correlations and, in particular,
of high momentum components by looking at the population of strength above $k_F$
using realistic momentum distributions.
Similarly, we have quantified the contribution of momenta beyond the Fermi surface
in the kinetic and total energies with the help of the Galistkii--Migdal--Koltum sum-rule
based on spectral functions obtained in the ladder approximation. The change in 
nature of high momentum components as the isospin asymmetry 
is modified leads to 
a substantial decrease of the kinetic component of the symmetry energy with
respect to the FFG. The tensor components of the NN interaction are largely responsible
for this effect. 
The results discussed here only include two-body forces, but we do not expect that 
the overall features will change much when 3BFs are included. 
For quantitative predictions in nuclei and neutron-star matter, however, 
the effect of three-nucleon interactions is clearly needed. Work is underway in this direction. 

\acknowledgments
We thank I. Vida\~na for useful and motivating discussions.
This work has been supported by Grants No. FIS2008-01661 (Spain), No. 2009-SGR1289 
and No. FI-DGR2011 
from Generalitat de Catalunya, 
a Marie Curie Intra European Fellowship within the 7$^{th}$ 
Framework programme and STFC grant ST/F012012, 
and by COMPSTAR, an ESF Research Networking Programme.

\bibliographystyle{apsrev4-1}
\bibliography{biblio}

\end{document}